\documentclass[pra, a4paper, aps, twocolumn, showpacs,superscriptaddress, groupedaddress]{revtex4}
\usepackage{subfig}
\usepackage{multirow}
\usepackage{array}
\usepackage{color}
\usepackage[dvips]{graphicx}
\usepackage{ae}
\usepackage[T1]{fontenc}
\usepackage[ansinew]{inputenc}
\usepackage{amsmath}
\usepackage{amssymb}
\usepackage{graphicx}
\usepackage{color}
\usepackage[colorlinks]{hyperref}
\usepackage{lscape}
\begin{document}
\title{Efficient Test to Demonstrate Genuine Three Particle Nonlocality}
\author{Kaushiki Mukherjee}
\email{kaushiki_mukherjee@rediffmail.com}
\affiliation{Department of Applied Mathematics,  University of Calcutta,  92,  A.P.C. Road,  Kolkata-700009, India.}
\author{Biswajit Paul}
\email{biswajitpaul4@gmail.com}
\affiliation{Department of Mathematics, St.Thomas' College of Engineering and Technology, 4, Diamond Harbour Road, Alipore, Kolkata-700023, India.}
\author{Debasis Sarkar}
\email{dsappmath@caluniv.ac.in}
\affiliation{Department of Applied Mathematics,  University of Calcutta,  92,  A.P.C. Road,  Kolkata-700009, India.}

\begin{abstract}
According to the studies of genuine tripartite nonlocality in discrete variable quantum systems conducted so far, Svetlichny inequality is considered as the best Bell-type inequality to detect genuine (three way) nonlocality  of pure tripartite genuine entangled states. In the present work, we have considered another Bell-type inequality (which has been reported as the $99$-th facet of $NS_2$ local polytope in (J.-D. Bancal, et.al.,Phys. Rev.A \textbf{88}, 014102 (2013)), to reveal genuine tripartite nonlocality of generalized GHZ(Greenberger-Horne-Zeilinger) class and a subclass of extended GHZ class states(\cite{ACN}) thereby proving the conjecture given by Bancal, et.al.\cite{BAL} for the GGHZ class and the subclass of extended GHZ states. We compare the violation of this inequality with Svetlichny inequality which reveals the efficiency of the former inequality over the latter to demonstrate genuine nonlocality using the above classes of quantum states. Even in some cases discord monogamy score can be used as a better measure of quantum correlation over Svetlichny inequality for those classes of pure states. Besides, the $99$-th facet inequality is found efficient not only for revealing genuine nonlocal behavior of correlations emerging in systems using pure entangled states but also in some cases of mixed entangled states over Svetlichny inequality and some well known measures of entanglement .
\end{abstract}
\date{\today}
\maketitle
\section{Motivation}
The correlated statistics arising by performing local measurements on an entangled state are nonlocal in nature in the sense that they violate a Bell inequality \cite{RRR,RLL}. Nonlocality has been a core part of quantum mechanics which has been supported by many experimental evidences\cite{GEN,ASP}. So far there has been a lot of analysis about the nonlocal nature of correlations arising between two space-like separated quantum systems exploring different tools demonstrating nonlocality in bipartite systems, such as, different forms of Bell-type inequalities, nonlocality witnesses, etc.  However, the understanding about multipartite nonlocality has not yet reached a satisfactory level due to increasing complexity while one shift from a bipartite to a multipartite scenario. In this context, perhaps the most interesting topic is that of genuine nonlocality. It may be referred to as correlations emerging in a $n$-party quantum system when all of the spatially separated parties constituting the system are nonlocally correlated. It is the strongest form of nonlocality. Apart from enriching quantum mechanics theoretically, study of nonlocality has significant contributions in the field of applications, for example, developing practical quantum information processing protocols \cite{SEE,HEI,CHE,SCA,SE1,ZOL,LUX,LAF,BOU,ROO}, communication complexity problems\cite{CLE}, device independent quantum cryptography \cite{BAR,MAS,ACD,MAP}, randomness expansion \cite{PIR,COL}, measurement-based quantum computation \cite{RAU,RAS}, etc. Besides, multipartite long-range correlations can play an important role in various condensed matter systems\cite{ROS,BRU}, quantum phase transitions \cite{ZOL,OST,VID,NIE}. Svetlichny laid the cornerstone in the study of genuine multiparty nonlocality.
\subsection{Svetlichny inequality} Svetlichny introduced a different form of correlations namely hybrid local-nonlocal form:
\begin{widetext}
\begin{equation}\label{1}
    P(abc/XYZ) =\sum_{\lambda}q_{\lambda}P_{\lambda}(ab/XY ) P_{\lambda}(c/Z)+\sum_{\mu} q_{\mu}P_{\mu}(ac/XZ) P_{\lambda}(b/Y ) +\sum_{\nu} q_{\nu}P_{\nu}(bc/Y Z) P_{\nu}(a/X);
\end{equation}
\end{widetext}
where $a,\,b,\,c\,\in\{0,1\}$ denote the outputs and $X,\,Y,\,Z\,\in\{0,1\}$ denote the inputs of the three parties Alice, Bob and Charlie respectively. Here $0\,\leq\,q_{\lambda},q_{\mu},q_{\nu}\,\leq\,1$ and $\sum_{\lambda}q_{\lambda}+\sum_{\mu}q_{\mu}+\sum_{\nu}q_{\nu}=1.$
The correlations which cannot be written in this form are genuine nonlocal (we will refer it as, \textit{$S_2$}(Svetlichny) nonlocal, in future). In order to detect genuine nonlocality, a Bell-type inequality $S\,\leq\,4$ is provided where
\begin{widetext}
\begin{equation}\label{2}
   S\,=\, \langle X_0Y_0Z_0\rangle+\langle X_1Y_0Z_0\rangle-\langle X_0Y_1Z_0\rangle+\langle X_1Y_1Z_0 \rangle +\langle X_0Y_0Z_1\rangle-\langle X_1Y_0Z_1\rangle+\langle X_0Y_1Z_1\rangle +\langle X_1Y_1Z_1\rangle ,
\end{equation}
\end{widetext}
violation of which guarantees \textit{Svetlichny} nonlocality(\cite{SVE}) ($S_2$ nonlocality(\cite{BAL}). Svetlichny inequality has been further generalized to arbitrary number of parties \cite{COI,SVI} and arbitray dimension \cite{BAN,CHI}. This inequality (\ref{2}) is violated by GHZ and W states \cite{GHO,AJO,LAV} and can be regarded as a unique witness of genuine tripartite nonlocality in a scenario where each of the three parties performs two dichotomic measurements \cite{SVE}. However, it can be regarded as a sufficient criteria only for detecting genuine nonlocality as there exist some genuine nonlocal correlations which satisfy this inequality \cite{BAL,GAL,GGO}.
\subsection{Incompleteness of $S_2$ correlations } The definition(\ref{1}) being general, no restrictions were imposed on the bipartite terms used in the inequality (\ref{2}). There may be one-way or both way signaling between a pair of parties or both the parties may perform simultaneous measurements. Thus, Svetlichny type nonlocality and hence the inequality (\ref{2}) lacks physical motivation which in turn may lead to grandfather-style paradoxes \cite{BAL} and inconsistency in operational purposes \cite{GAL,GGO}.  In \cite{BAL} Bancal et.al., removed this sort of ambiguity by putting restrictions on the bipartite terms and thereby provided two alternative definitions of nonlocality: $NS_2$(no signaling) type and $T_2$(time ordering) type.
 \subsection{$NS_2$ correlations and Bancal etal.'s conjecture} Among these two types, in the $NS_2$ type nonlocality, bipartite correlation terms abide by the no signaling criteria:
\begin{widetext}
\begin{equation}\label{3}
    P(abc/XY Z) =\sum_{\lambda}q_{\lambda}P_{\lambda}(ab/XY ) P_{\lambda}(c/Z)+\sum_{\mu} q_{\mu}P_{\mu}(ac/XZ) P_{\lambda}(b/Y ) +\sum_{\nu} q_{\nu}P_{\nu}(bc/Y Z) P_{\nu}(a/X)
\end{equation}
\end{widetext}
where the bipartite terms satisfy no signaling conditions of the form:
\begin{equation}\label{4i}
    P_{\lambda}(a/XY ) = P_{\lambda}(a/XY')\, ~~\forall\, a,\, X,\, Y,\, Y'
\end{equation}
\begin{equation}\label{4ii}
   P_{\lambda}(b/XY) = P_{\lambda}(b/X'Y) \, ~~\forall \,b, X,\, X',\, Y.
\end{equation}

The correlations which are of the above form are called $NS_2$(no signaling) local. Otherwise, they are $NS_2$ nonlocal. The no signaling constraints imposed on the bipartite terms exclude the possibility of the outcomes of one or two parties influencing the inputs of the remaining one. Hence this form of nonlocality is in general weaker than that of Svetlichny nonlocality. In \cite{BAL}, $185$ Bell-type inequalities are given which constitute the full class of facets of $NS_2$ local polytope. Violation of any of these facets (Bell-type inequalities) guarantees $NS_2$ nonlocality. Svetlichny inequality constitute the $185$-th class.  So far, mostly the $185$-th class facet inequality, i.e., Svetlichny inequality(\ref{2}) has been used as a tool to demonstrate genuine nonlocality. For instance, in \cite{GHO} S. Ghose et.al., analyzed the relation between genuine nonlocality and three-tangle (genuine tripartite entanglement measure)\cite{COF} for the class of GGHZ \cite{ACN,DUR} and Maximal Slice(MS) states \cite{CAR}.The latter are included in the subclass $S$ of extended GHZ states\cite{ACN,DUR}). They showed that up to projective measurements, GGHZ states show genuine nonlocality for three-tangle ($\tau$) greater than $\frac{1}{3}$ whereas MS states violate Svetlichny inequality for any positive value of $\tau$. In \cite{AJO}, closed forms of the bounds of Svetlichny inequality were derived for extended GHZ and W class of states. In the framework where  a complete set of Bell-type inequalities were introduced by Bancal et. al.(\cite{BAL}), it was conjectured that genuine tripartite entanglement of a pure state guarantees genuine nonlocality. With the help of one of these Bell-type inequalities ($168$-th class), Adesso et. al., investigated the genuine tripartite nonlocality of three-mode Gaussian states in continuous variable systems \cite{ADE}. Besides, in\cite{ALM} Almeida et.al., provided sufficient criteria for a quantum system to be fully nonlocal according to a given partition and also to be genuinely multipartite nonlocal. Other than the statistical approach \cite{SVE,BAL} genuine nonlocality of tripartite states was revealed with the aid of Hardy type argument \cite{RAH,CHN,SYU}. Apart from demonstrating nonlocality in quantum systems, an information theoretic measure of discord monogamy score \cite{RPR,PRA} has been used to exploit quantum correlations beyond entanglement for multipartite systems. Although all these studies have contributed in providing different means of demonstrating nonlocality and thereby quantumness in different physical systems, yet, it is still a matter of interest to develop better tools for exploiting the same. Our work basically focuses in this direction.
 \subsection{Summary of the work} We have used one of the facets of $NS_2$ local polytope as a better tool than Svetlichny for demonstrating three way nonlocal nature of correlations emerging in discrete variable quantum systems. This in turn is useful to analyze the existing relation in between nonlocal nature of the resulting correlations with that of genuine tripartite entanglement of the quantum states involved in the systems. We have used the $99$-th class facet of the $NS_2$ local polytope \cite{BAL}. The $99$-th facet inequality (\cite{BAL}) is given by:
\begin{equation}\label{5}
 NS \leq 3.
\end{equation}
where $NS = \langle X_1Y_1\rangle +   \langle X_0Y_0Z_0\rangle +   \langle Y_1Z_0\rangle +   \langle X_1Z_1\rangle -  \langle X_0Y_0Z_1\rangle.$ For this facet, we derive the closed form of maximum violation for two class of pure tripartite states: GGHZ and a subclass ($S$) of extended GHZ states (\cite{ACN}) under projective measurements. Interestingly the $99$-th class facet (Eq.\ref{5}) helps us to reveal genuine nonlocality of GGHZ states for any non zero value of $\tau$ unlike Eq(\ref{2}), where the same is  revealed only for $\tau >\frac{1}{3}$. This in turn proves the conjecture made by Bancal et.al.,(\cite{BAL}) for the GGHZ class of states. Further, a comparative study of the violations of $99$-th facet inequality and Svetlichny inequality for the subclass $S$ of extended GHZ states reveals that  the former(\ref{5}) gives advantage over the latter for a certain range of $\tau$. In particular, MS class of states (a subclass of $S$) exhibits genuine tripartite nonlocality for any positive amount of $\tau$ for both of these facet inequalities. Consequently, $99$-th class facet inequality(\ref{5}) emerges as a more efficient tool compared to Svetlichny for revealing three-way nonlocality in a quantum system using the above mentioned two classes of pure tripartite states. Quantumness is revealed in a system via generation of nonlocal correlations. From that perspective $99$-th facet inequality serves as a good measure of quantum correlation behaving in a similar manner as discord monogamy score ($\delta_D$) for the complete GGHZ class and also for subclass $S$ of extended GHZ class. The similarity in the pattern of variation of discord monogamy score ($\delta_D$) and that of amount of violation of $99$-th facet inequality with the state parameter or amount of entanglement also supports the fact that $\delta_D$ helps us to detect genuine nonlocality via revelation of quantumness even in some cases where Svetlichny inequality cannot be used as a tool to detect genuine nonlocality of these two classes of pure states. For instance, considering GGHZ class of states, $\delta_D>0$ in the subinterval $[0,0.393]$ of the interval $[0,\frac{\pi}{4}]$ of the state parameter ($\eta$) where the class reveals nonlocality (guaranteed by violation of the $99$-th facet inequality) but Svetlichny cannot be used to demonstrate the same. Similar sort of results also exist for subclass($S$) of extended GHZ class of states. In case of mixed tripartite states, the main focus so far was on analyzing the entanglement properties of the states via various entanglement measures. Here we have studied  nonlocal behavior of noisy GHZ states, noisy MS states and some family of high rank mixed states(whose explicit expressions for the three-tangle($\tau$) are reported in \cite{LOH,ELT,EJU,JUN,SJH}) by providing closed forms of violation of $99$th facet inequality(\ref{5}). Our work can be organized as follows: Section (II) deals with some mathematical prerequisites.  Bounds for violation of the $99$th facet(\ref{5}) are discussed in Section(III). Next three sections(IV-VI) deal with various sides of getting advantage of the $99$th facet over Svetlichny inequality(\ref{2}) from physical point of view. In Section(VII) we end up with a conclusion.
\section{Preliminaries}
To begin with we first explain the three-tangle ($\tau$) which is used as the measure of genuine tripartite entanglement \cite{COF}:
\begin{equation}\label{5a}
    \tau\,=\,\mathcal{C}^2_{1(23)}\,-\,\mathcal{C}^2_{12}\,-\,\mathcal{C}^2_{13}
\end{equation}
$\mathcal{C}^2_{1(23)}$ is a measure of entanglement between the first qubit($1$) and
the joint state of the last two qubits ($2,3$). $\mathcal{C}^2_{12}$ and $\mathcal{C}^2_{13}$ are the concurrences \cite{WOO} measuring nature of bipartite entanglement between qubits $1$, $2$ and $1$, $3$ respectively. $\tau$ remains invariant under permutation of the indices ($1,2,3$) and lies in the interval $[0,1]$. For separable and bi-separable states $\tau\,=\,0$ and for GHZ state, it is $1$. For a three-qubit pure state of the form:
\begin{equation}
|\psi \rangle\,=\,a|011\rangle \,+\,b|101\rangle \,+\, c|110\rangle \,+\,d|000\rangle \,+\,he^{i\gamma}|111\rangle
\end{equation}
three-tangle is given by
\begin{equation}\label{6a}
\tau\,=\,4d\sqrt{(dh^2-4abc)^2+16abcdh^2cos^2\gamma}
\end{equation}

\textit{Quantum Monogamy Score:} Monogamy constrains the sharing of quantum correlations among subsystems of a multipartite quantum state. For a tripartite quantum state $\rho_{ABC}$, a bipartite quantum correlation measure ($\mathcal{Q}$) is monogamous (with $A$ as the ``nodal observer``) if
\begin{equation}\label{6b}
    \mathcal{Q}(\rho_{A:BC})\geq \mathcal{Q}(\rho_{AB})+\mathcal{Q}(\rho_{AC}).
\end{equation}
Here, $\mathcal{Q}(\rho_{AB})$, $\mathcal{Q}(\rho_{AC})$ and $ \mathcal{Q}(\rho_{A:BC})$ denotes quantum correlation (considering $\mathcal{Q}$ as  a bipartite measure) between subsystems $(A,B)$, $(A,C)$ and between system $A$ and subsystems $B$ and $C$ taken together respectively. The above inequality can be recast as
\begin{equation}\label{6i}
    \delta_{\mathcal{Q}}\equiv  \mathcal{Q}(\rho_{A:BC})-\mathcal{Q}(\rho_{AB})-\mathcal{Q}(\rho_{AC})\geq\,0.
\end{equation}
$\delta_{\mathcal{Q}}$ is regarded as the monogamy score. If discord ($D$) is considered as the quantum correlation measure ($\mathcal{Q}$) then we get discord monogamy score \cite{RPR,PRA}:
\begin{equation}\label{6ii}
    \delta_D\,=\,D(\rho_{A:BC})\,-\,D(\rho_{AB})\,-\,D(\rho_{AC})
\end{equation}
where $D(\rho_{AB})$ denotes quantum discord between subsystems $A$ and $B$ \cite{HEN,OLI,MOD}. Discord monogamy score ($\delta_D$) has been interpreted as a  multiparty information-theoretic  quantum correlation measure \cite{PRA}.  A tripartite state $\rho_{ABC}$ satisfies monogamy relation (Eq.(\ref{6b})), if $\delta_D>0,$ whereas it violates Eq.\ref{6b}, if $\delta_D\leq0$.\\
\section{Bounds for Bell inequality violation}
\subsection{GGHZ class}
First we consider generalized Greenberger-Horne-Zeilinger (GGHZ) class of states \cite{ACN}:
\begin{equation}\label{6}
  |\varphi_{GGHZ}\rangle\,=\,  \cos\eta |000\rangle\,+\sin\eta |111\rangle\,\,\textmd{where}\, \eta\in[0,\,\frac{\pi}{4}].
\end{equation}
The GHZ states(\cite{ZUK1}) are for $\eta=\frac{\pi}{4}$. The GGHZ states are significant for the study of nonlocality as they are the only class of pure three qubit states such that all information about them can be coded in the corresponding tripartite correlations (they cannot be uniquely reconstructed from the corresponding reduced two-qubit states) \cite{LIN}. Besides, because of inherent symmetry in their structure \cite{CAR}, they possesses some unique entanglement characteristics for which they are useful in various information processing protocols.
For this class of states the bound of Eq.(\ref{5}) is given by(see Appendix.A),
\begin{equation}\label{7}
    S\,\leq\, B_1 \,\textmd{where}\,\,B_1=1\,+\,2\sqrt{1+\sin^22\eta}.
\end{equation}
Here $\tau\,=\,\sin^22\eta$. Except for $\eta =0 $, the class of states (\ref{6}) has genuine tripartite entanglement. Thus, from Eq.(\ref{7}) we have:
\begin{equation}\label{8}
   B_1\,=\, 1+2\sqrt{1+\tau}.
\end{equation}
The closed form of violation of Svetlichny inequality is given by (\cite{GHO}) $S<B_2$ where
\begin{equation*}
 B_2 =4\sqrt{1-\tau}\,~~\textmd{if}\,\tau\leq\frac{1}{3}
\end{equation*}
\begin{equation}\label{8i}
 = 4\sqrt{2\tau}\,~~\textmd{if}\,\tau\geq\frac{1}{3}.
\end{equation}
Hence for any amount of genuine tripartite entanglement ($\tau > 0$), genuine nonlocality is revealed (using $99$-th facet inequality), i.e., $B_1 > 3 $ (FIG.2). Whereas the restrictions imposed on the amount of tripartite entanglement is $\tau\in(\frac{1}{3},1]$ to show genuine nonlocality (FIG.1) when Svetlichny inequality is used \cite{GHO}.
\begin{figure}[htb]
	\begin{center}
		\includegraphics[width=3.3in]{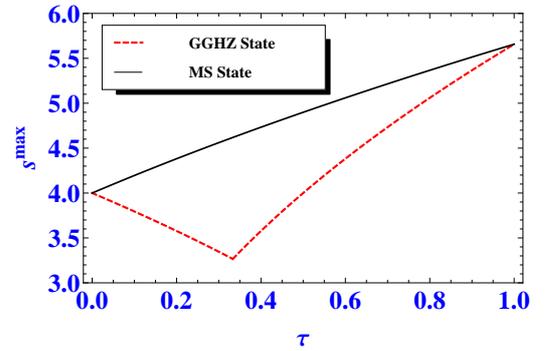}
		\caption{\emph{ The variation of violation of Svetlichny inequality with three-tangle $\tau$ is plotted for GGHZ (dashed line) and MS (solid line) subclasses (\cite{GHO}) of states. MS violates Svetlichny for any value of $\tau$ unlike that of GGHZ class which reveals genuine nonlocality ($S_2$ nonlocality) for $\tau > \frac{1}{3}.$}}
	\end{center}
\end{figure}

\begin{figure}[htb]
	\centering
	\includegraphics[width=2.8in]{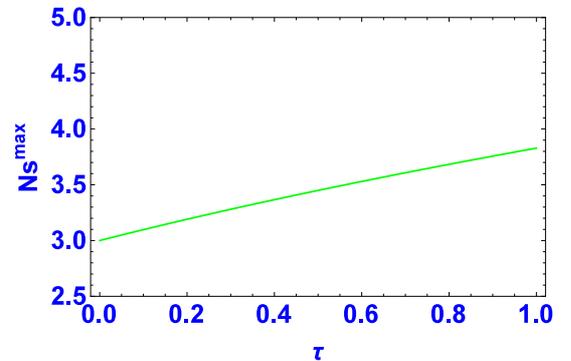}
	\caption{\emph{ The amount of nonlocality in terms of the $99$-th class facet inequality violation ($NS_2$-nonlocality (\cite{BAL}) increases monotonically with the amount of tripartite entanglement ($\tau$) for both GGHZ and MS subclasses. This figure shows that violation is obtained for any amount of three-tangle.}}
\end{figure}

\subsubsection{Bell inequality violation as a measure of quantumness}
It has already been mentioned that as violation of a Bell inequality gurantees nonlocal behaviour of a physical system, hence it may be considered as a mean of detecting quantumness(nonclassicality) of a quantum state. So from that perspective violation of a Bell inequality may be compared with a standard measure of quantumness. For that purpose we have considered here discord monogamy score. For GGHZ class, discord monogamy score(see Appendix.C) is given by:
\begin{equation}\label{8ii}
    \delta_D\,=\,-(\cos^2\eta\log_2(\cos^2\eta) + \sin^2\eta\log_2(\sin^2\eta))
\end{equation}

Both $B_1$ and $\delta_D$ are functions of state parameter $\eta$. Figure 3 shows that the curve of the quantum correlation measure ($\delta_D$) and that of the bound ($B_1$) both vary in a similar fashion. For $\eta\in[0,\frac{\pi}{4}]$ Svetlichny inequality (\ref{2}) is not violated in the sub interval $[0,0.393]$ whereas states show nonlocal character  by violating $99$-th facet inequality (\ref{5}). Also, within this sub interval the states have positive discord monogamy score ($\delta_D$). The states having higher measure of quantum correlation show greater violation of Eq.(\ref{5}) unlike that of violation of Svetlichny inequality (\ref{2}) (FIG.3).
\begin{figure}[htb]
\centering
\includegraphics[width=2.8in]{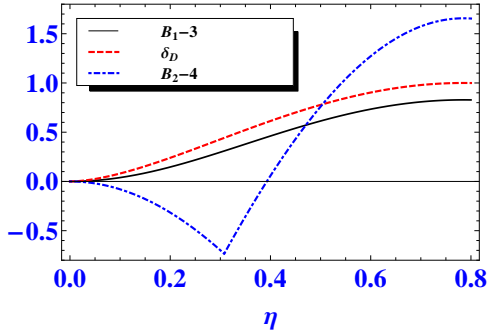}
\caption{\emph{ The degree of violation of $99$-th facet inequality (\ref{5}), Svetlichny inequality (\ref{2}) and discord monogamy score ($\delta_D$) are plotted against state parameter $\eta$. Clearly,  the nature of the curve representing degree of violation (\ref{5}) is the same as that of $\delta_D$ (\ref{8ii}) curve which in turn implies that the degree of violation of Eq.(\ref{5}) is more for the states having greater value of $\delta_D$. However, if violation of Eq.(\ref{2}) is considered, there exist states belonging to the GGHZ class for which $\delta_D>0$ but those states cannot reveal genuine nonlocality in terms of violation of Svetlichny inequality (\ref{2}).}}
\end{figure}
\subsection{Extended GHZ class}
The subclass ($S$) of extended GHZ class of states has the form(\cite{ACN}):
\begin{equation}\label{9}
  |\chi\rangle\,=\,  \lambda_0 |000\rangle\,+\, \lambda_3|110\rangle\,+\,\lambda_4|111\rangle\,
\end{equation}
where, $\lambda_i\in[-1,1],$ and $\sum_{i=0,3,4} \lambda_i^2=1.$

For $\lambda_0\,=\,\frac{1}{\sqrt{2}}$, $\lambda_3\,=\,\frac{\cos\eta}{\sqrt{2}}$ and $\lambda_4\,=\,\frac{\sin\eta}{\sqrt{2}},$ we get the\textit{ MS states}(\cite{CAR}):
\begin{equation}\label{10}
    \frac{1}{\sqrt{2}}(|000\rangle\,+\, \cos\eta|110\rangle\,+\,\sin\eta|111\rangle))\,\,\eta\in[0,\frac{\pi}{4}].
\end{equation}
\subsubsection{Bounds for MS subclass}
The MS subclass of extended GHZ states have various practical applications. For instance the maximally entangled GHZ state: $\frac{ |000\rangle\,+\, |111\rangle}{\sqrt{2}}$ (belonging to both MS and GGHZ subclasses) has been used in various physical processes \cite{LAF,BOU,ROO,NEL}. For MS state, the $NS$ bound is given by (see Appendix-B):
\begin{equation}\label{11}
    B_3\,=\,1\,+\,2\sqrt{1+\sin^2\eta}.
\end{equation}

Here, $\tau\,=\,\sin^2\eta.$ Using this relation, the above bound gets modified as: $$ B_3\,=\, 1+2\sqrt{1+\tau}.$$
Hence $B_3$ is of the same form as that of $B_1$ (\ref{8}). Another class of MS states obtained by swapping qubits of second and third parties yield the same bound ($B_3$). So any amount of genuine tripartite entanglement suffices to produce violation of Eq.(\ref{5}). A comparison of the different values of the bounds ($B_1$, $B_3$) obtained by using GGHZ and MS class suggest that for $\sin^2\eta>0.75$ GGHZ class yields more nonlocality in terms of violation of Eq.(\ref{5}). Besides, for a given amount of genuine tripartite entanglement both the classes give the same amount of violation of Eq.(\ref{5}) in contrast to the violation of Svetlichny inequality where MS gives more violation than GGHZ \cite{GHO,EMA,SOH}. It is also interesting to note that variation of maximum quantum violation of $99$-th facet inequality (upto projective measurements) by a tripartite state belonging to GGHZ or MS class  with that of tripartite entanglement measure $\tau$ is analogous to variation of maximum quantum violation of CHSH inequality \cite{CLA} ($2\sqrt{1+C_{12}^2}$) for two-qubit pure states \cite{GIS,POP} with that of bipartite entanglement measure $C_{12}^2$ (FIG.4).
\begin{figure}[htb]
	\centering
	\includegraphics[width=2.8in]{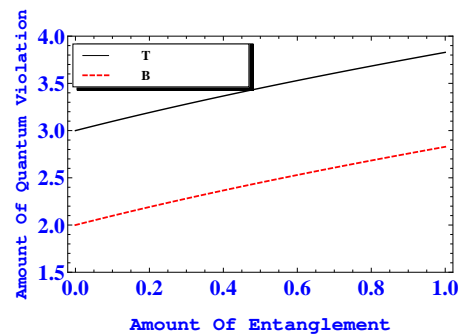}
	\caption{\emph{ The red curve gives variation of maximum quantum violation($T$) of $99$-th facet inequality (\ref{5}) with that of $\tau$ whereas the dashed curve represents variation of maximum quantum violation($B$) of CHSH inequality with $C_{12}^2$. Evidently, both the violations vary with the corresponding entanglement measures in a similar fashion.}}
\end{figure}
\subsubsection{Bounds for subclass $S$ }
 The amount of tripartite entanglement of the subclass $S$ of extended GHZ states (\ref{9}) is: $\tau\,=\,4\lambda^2_0\lambda^2_4$ and the measure of bipartite entanglements are: $C_{12}^2=4\lambda^2_3\lambda^2_0$, $C_{13}^2=0$ and $C_{23}^2=0$.
The bound for Svetlichny inequality is (\cite{AJO}): $$ S\,\leq\, B_4$$ where
\begin{equation*}
  B_4\,=\,4\sqrt{1\,-\,\tau},~~\textmd{if}\, ~\tau\,\leq\frac{1-C_{12}^2}{3}
\end{equation*}
\begin{equation}\label{12}
  \,\,\,=\,4\sqrt{C_{12}^2+2\tau},~~\textmd{if}\, ~\tau\,>\frac{1-C_{12}^2}{3}.
\end{equation}
and for the $99$-th facet inequality (\ref{5}) the bound (see Appendix-B) is: $$S\,\leq\, B_5$$ where
\begin{equation*}
    B_5\,=\,3\,~~\textmd{if}\,~\tau\,\leq\frac{C_{12}^2(1-C_{12}^2)}{1+C_{12}^2},
\end{equation*}
\begin{equation}\label{13}
 \qquad \,\,=\,1\,+\,\sqrt{A+2C}+\sqrt{A-2C}\,\,~\textmd{if}\,~\tau\,>\frac{C_{12}^2(1-C_{12}^2)}{1+C_{12}^2}
  \end{equation}

where, $ A= 1+\tau,\,\,C=\sqrt{C_{12}^2(1-\tau-C_{12}^2)}.$  As, $4\sqrt{1\,-\,\tau}\,\leq\,4$, this class violates Svetlichny inequality (\ref{2}) if $C_{12}^2+2\tau > 1$. Hence up to projective measurements, $S_2$ nonlocality is revealed in a quantum system using this subclass ($S$) only if $\tau\in(\frac{1-C_{12}^2}{2}\,,\,1].$ But for the states belonging to $S$  for which $\lambda_0 = \frac{1}{\sqrt{2}}$, $S_2$ nonlocality is obtained for any positive value of $\tau$(See Appendix D). As the Maximal Slice states (\ref{10}) belongs to this category, the results related to nonlocality of MS states reported in \cite{GHO} can be easily recovered from here. Now, $99$-th facet inequality is violated if $B_5 > 3,$ which in turn implies that $\tau$ must lie in $(\frac{C_{12}^2(1-C_{12}^2)}{1+C_{12}^2},\,1]$. Clearly any tripartite state belonging to the subclass S and characterized by $\lambda_0 = \frac{1}{\sqrt{2}}$(See Appendix D) violates $99$-th facet inequality(\ref{5}). But in general for any state belonging to the subclass $S$, unlike Svetlichny inequality (\ref{2}), the $99$-th facet inequality (\ref{5}) helps us to reveal $NS_2$ nonlocality of the correlations emerging from the quantum systems using the states of the subclass $S$ of extended GHZ class whose tripartite entanglement is characterized by $\tau$ lying in the interval $(\frac{C_{12}^2(1-C_{12}^2)}{1+C_{12}^2},\,\frac{1-C_{12}^2}{2}]$ (FIG.5). Clearly, for any amount of bipartite entanglement $C_{12}^2$ there is a region of advantage of $99$-th facet inequality (\ref{5}) over Svetlichny inequality (\ref{2}).
 \begin{figure}[htb]
 	\centering
 	\includegraphics[width=2.8in]{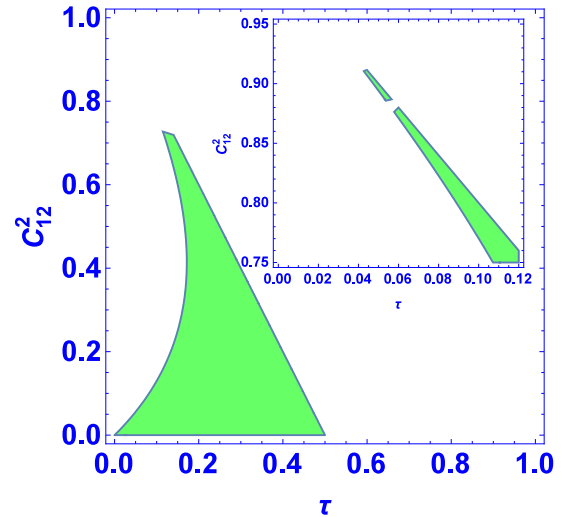}
 	\caption{\emph{ The shaded region indicates the area where $99$-th facet inequality (\ref{5}) emerges as a more efficient tool over Svetlichny inequality (\ref{2}) for revealing nonlocal nature of tripartite correlations in a quantum system using any state of the subclass $S$ of the extended GHZ states. Clearly, the region shrinks as bipartite entanglement ($C_{12}^2$) increases. Only small portion of advantage is obtained as $C_{12}^2$ increases above $0.75$ which is shown separately in the figure.}}
 \end{figure}
\subsubsection{Measure of quantumness}
The discord monogamy score ($\delta_D$) of  $|\chi\rangle$ (see Appendix C)is:
 \begin{equation}\label{13i}
    -\frac{((1 - \sqrt{1 - \tau}) \ln(\frac{1 - \sqrt{1 - \tau}}{2})\,+\,(1 +\sqrt{1 - \tau}) \ln(\frac{1 + \sqrt{1 - \tau}}{2}))}{\ln(4)}.
 \end{equation}

Quantum correlation measure ($\delta_D$) increases monotonically (FIG.6) with that of $\tau$. The degree of violation of $99$-th facet inequality (\ref{5}) varies in a similar fashion like discord monogamy score ($\delta_D$)(for different fixed values of the bipartite entanglement measure $C_{12}^2$) unlike that of the violation of Svetlichny inequality (\ref{2}) where the curve shows no such fixed behavior.
 \begin{widetext}
 	\begin{center}
 		\begin{figure}
 			\begin{tabular}{|c|c|}
 				\hline
 				\subfloat[$C_{12}^2\,=\,0.1$]{\includegraphics[trim = 0mm 0mm 0mm 0mm,clip,scale=0.65]{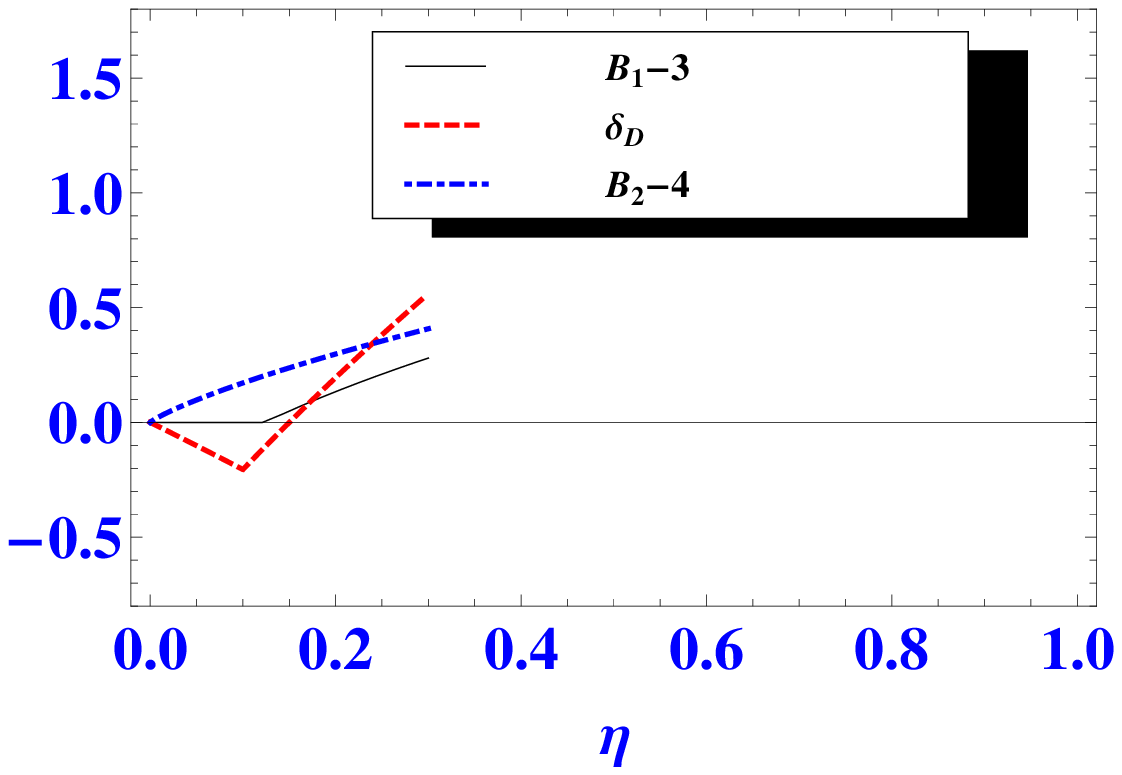}} &
 				\subfloat[$C_{12}^2\,=\,0.3$]{\includegraphics[trim = 0mm 0mm 0mm 0mm,clip,scale=0.65]{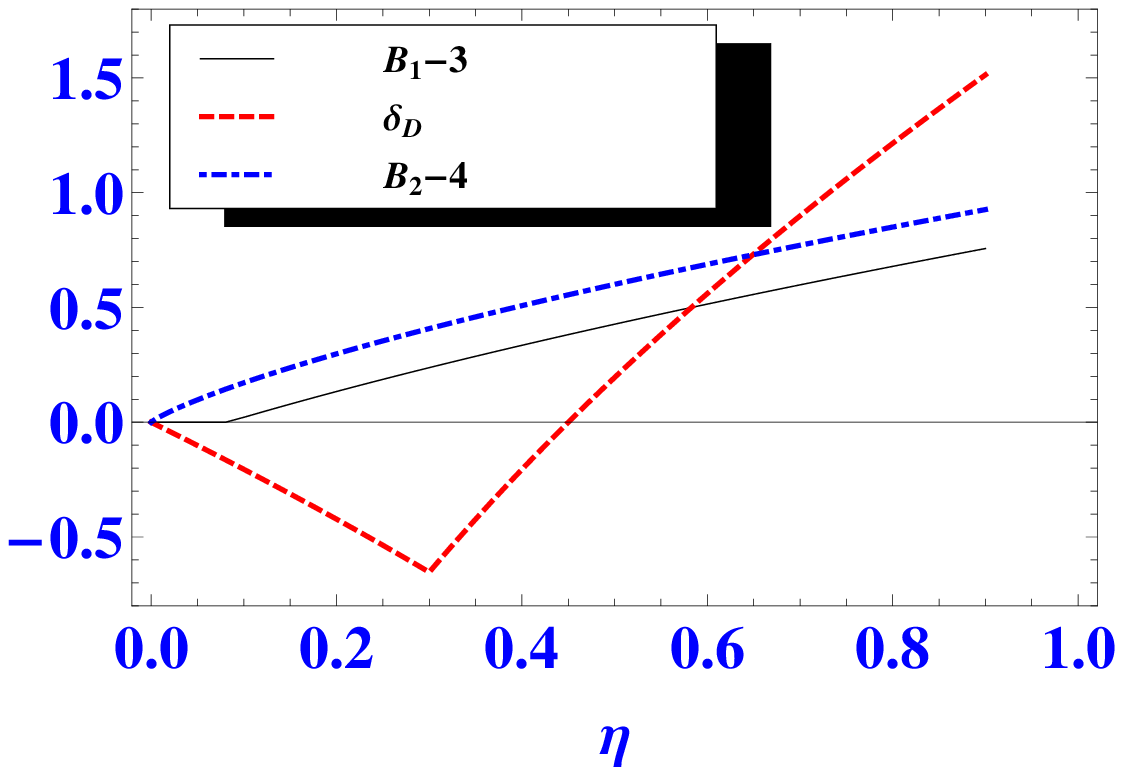}}\\
 				\hline
 				\hline
 				\subfloat[$C_{12}^2\,=\,0.5$]{\includegraphics[trim = 0mm 0mm 0mm 0mm,clip,scale=0.65]{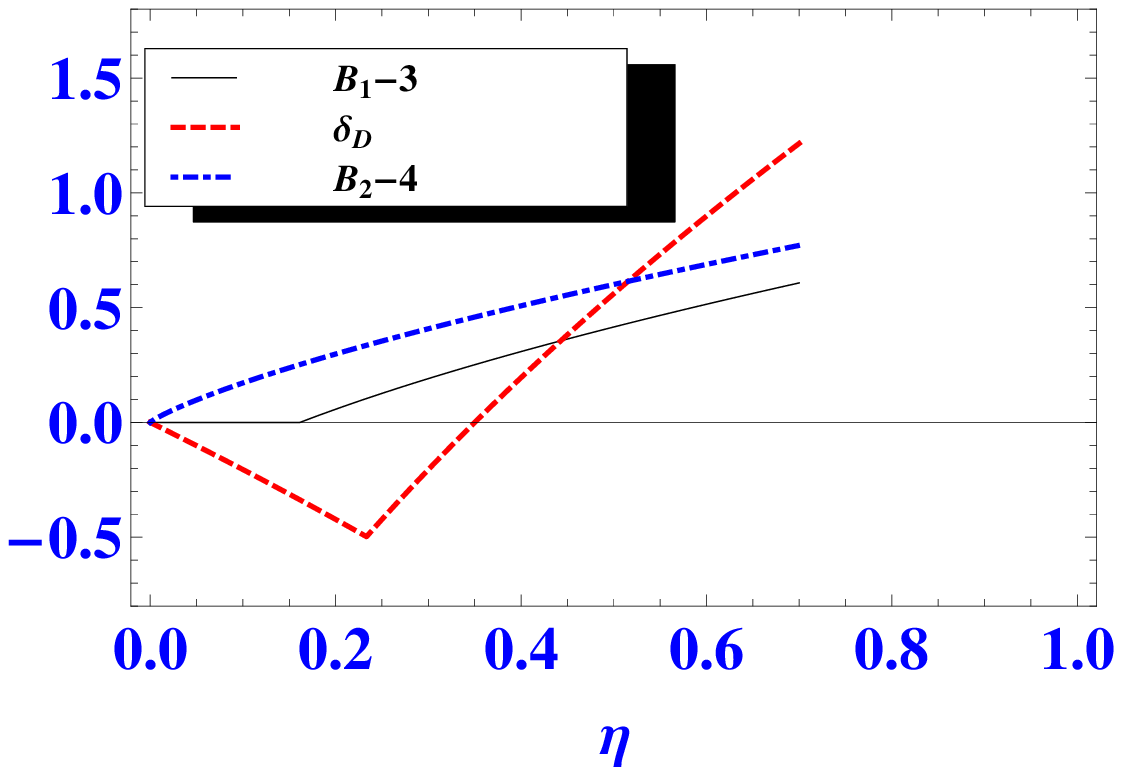}} &
 				\subfloat[$C_{12}^2\,=\,0.7$]{\includegraphics[trim = 0mm 0mm 0mm 0mm,clip,scale=0.65]{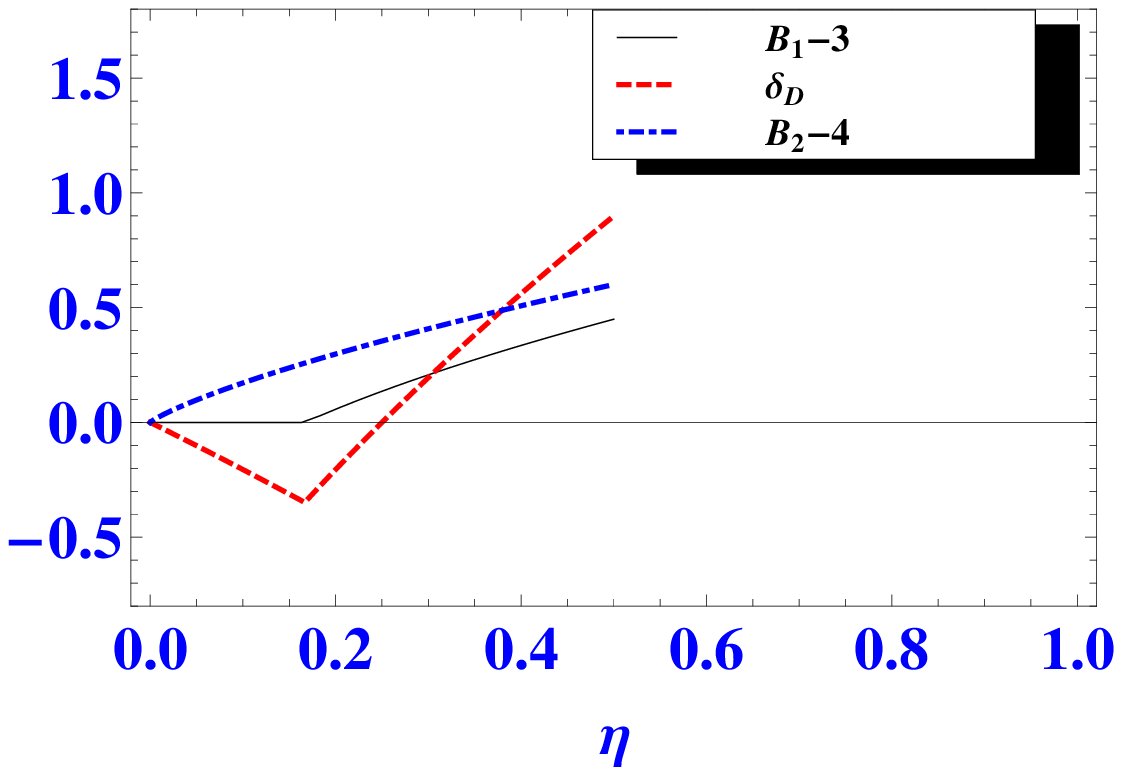}} \\
 				\hline
 			\end{tabular}
 			\caption{\emph{ The degree of violation of $99$-th facet inequality (\ref{5}), Svetlichny inequality (\ref{2}) and discord monogamy score ($\delta_D$) for subclass $S$ of extended GHZ class are plotted against $\tau$ for different fixed values of $C_{12}^2$. For $\tau$ lying in the range indicated by portion of the $\tau$ axis intercepted in between the points where the blue curve and red curve cut the $\tau$ axis (other than the origin), nonlocal feature of the corresponding correlations is revealed via violation of $99$-th facet inequality (\ref{5}). For this range of $\tau$, discord monogamy score ($\delta_D$) serves as a better tool for detecting genuine nonlocality compared to Svetlichny inequality (\ref{2}).}}
 		\end{figure}
 	\end{center}
 \end{widetext}
 
\section{Resistance To Noise} An interesting observation from both theoretical and physical view point is that of resistance offered to noise by a pure entangled state. For that, we consider a source $\mathcal{S}$ producing a noisy three qubit state(obtained after passing the state through a depolarizing channel\cite{NIE,BEN,BNN,BRS}):
\begin{equation}\label{x4iii}
\omega=\alpha \rho+(1-\alpha)\frac{\mathbf{I}}{8},\,(\textmd{where}\,\alpha\in[0,1]).
\end{equation}
Here $\alpha$ denotes the visibility of three qubit state $\rho$ which is a measure of the resistance to noise. $1-\alpha$ denotes the probability with which white noise is introduced in the system by the source. The largest visibility ($\alpha$) for which $\omega$ is local is referred to as the \textit{local visibility threshold} ($V$). Considering $\rho\,=\,|\varphi_{GGHZ}\rangle\langle \varphi_{GGHZ}|$, using $99$-th facet inequality one can find $\alpha\in(\frac{3}{1\,+\,2\sqrt{1+\tau}}\,,1]$ whereas for Svetlichny (\cite{GHO}) the range is given by $\alpha\in(\frac{1}{\sqrt{2\tau}}\,,1]$. Hence from the point of lowering down the local visibility threshold ($V$), the efficiency of $99$-th facet inequality and Svetlichny (\ref{2}) depends on the amount of tripartite entanglement present in the corresponding state (produced by the source $S$) belonging to the GGHZ class (FIG.7). For $\rho\,=\,|\chi\rangle\langle\chi|$, a similar analysis (FIG.8) can be made as $99$-th facet inequality (\ref{5}) is violated for $\alpha\in(\frac{3}{1\,+\,\sqrt{A+2C}+\sqrt{A-2C}},\,1],$ whereas $\alpha\in(\frac{1}{\sqrt{2\tau+C_{12}^2}},\,1], $ if violation of Svetlichny inequality(\ref{2}) is considered.
\begin{widetext}
	\begin{figure}[htb]
		\centering
		\includegraphics[width=2.6in]{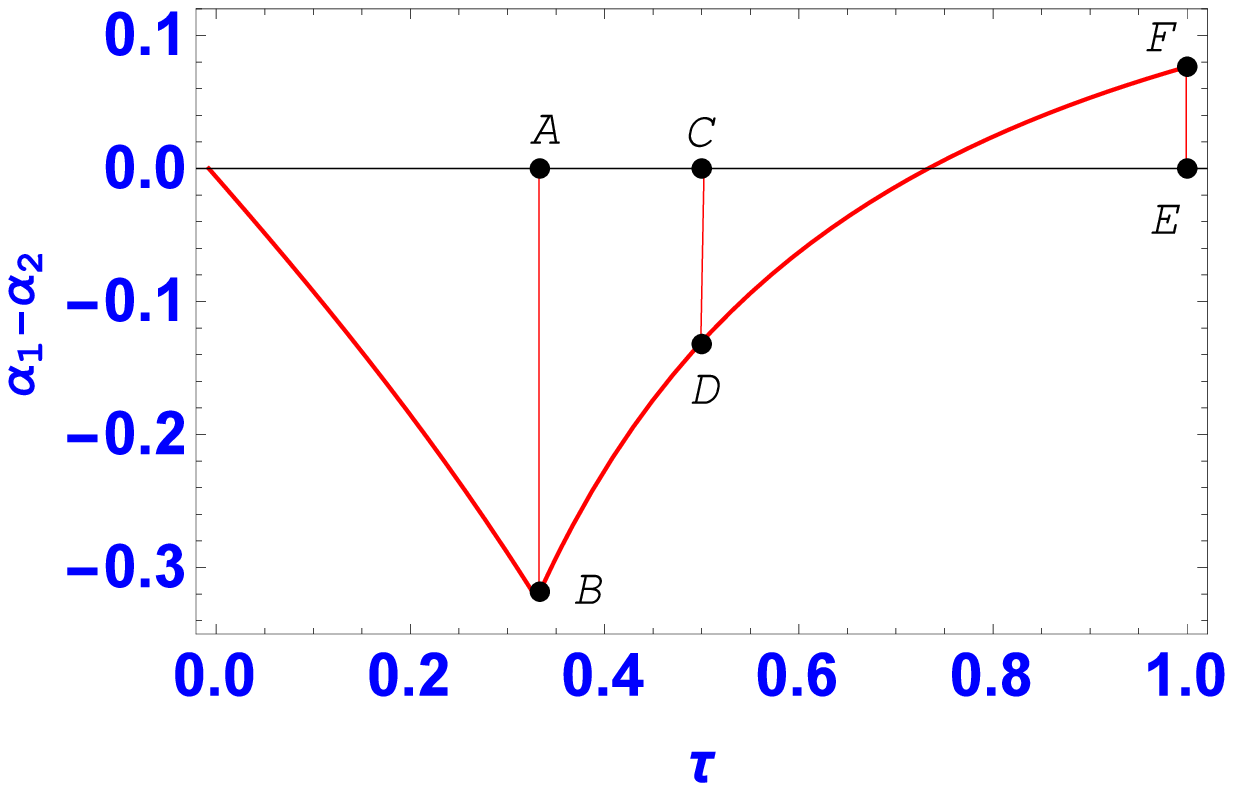}
		\caption{\emph{ The difference between the lower bounds $\alpha_1$ and $\alpha_2$ of the range of visibility ($\alpha$) of $\omega$ corresponding to the violation of $99$-th facet inequality (\ref{5}) and Svetlichny inequality (\ref{2}) respectively, is plotted against tripartite entanglement measure ($\tau$) of GGHZ class. $99$-th facet inequality gives maximum advantage in terms of lowering down local visibility threshold ($V$) for the noisy GGHZ state whose corresponding pure counterpart has tripartite entanglement ($\tau$) given by the point $A$. On the contrary Svetlichny inequality does the same for the noisy state corresponding to the point $E$. $99$-th facet inequality helps in lowering down local visibility ($V$) of those noisy GGHZ states whose corresponding pure counterparts are characterized by having $\tau$ corresponding to any point $D$ lying on the portion of the curve below $\tau$ axis.}}
	\end{figure}
\end{widetext}
\begin{widetext}
	\begin{figure}[htb]
		\centering
		\includegraphics[width=2.6in]{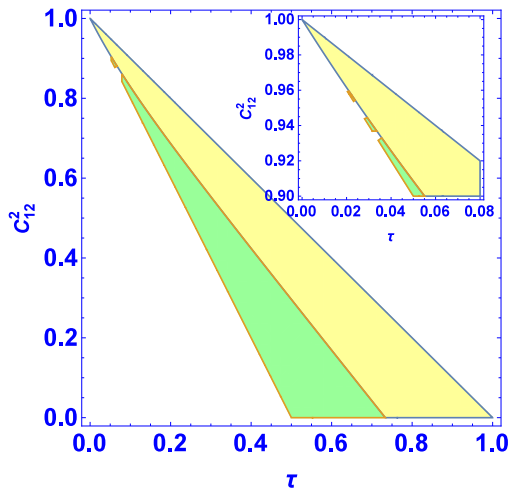}
		\caption{\emph{ For any value of $\tau$ and $C_{12}^2$ lying in the shaded region ($R$), $99$-th facet inequality (\ref{5}) gives advantage over Svetlichny inequality (\ref{2}), i.e., any pure entangled state (belonging to the subclass $S$ of extended GHZ class), characterized by its tripartite entanglement ($\tau$) and bipartite entanglement content ($C_{12}^2$) corresponding to a point lying in the region ($R$), is more resistant to noise when Eq.(\ref{5}) is considered. On the contrary, for any point lying in the region $R'$ the corresponding pure entangled state offers more resistance to noise via Svetlichny inequality (\ref{2}) violation. For any value of $C_{12}^2$ above $0.90$ discrete portions of $R$ and $R'$ are obtained (highlighted separately).}}
	\end{figure}
\end{widetext}
\subsection{ Exposure of individual qubits to noise}
Apart from considering a one parameter noisy state, obtained due to the effect of a depolarizing channel on a pure state, it is also interesting to deal with nonlocal character of the states when each of the three subsystems are subjected to:
\begin{itemize}
  \item independent depolarization with different strength
  \item individually exposed to amplitude damping with varying measure of dampness.
\end{itemize}
\subsubsection{Independent depolarization}
 If all the three subsystems of any  pure state of the GGHZ class(\ref{9}) are subjected to depolarization of varying strength then the noisy state is given by:
 \begin{widetext}
 \begin{equation}\label{r1}
   \frac{((J_1 +J_2)\cos^2\eta +
  2J_2 \sin^2\eta) |000\rangle\langle000|+((J_1 +J_2)\sin^2\eta +
  2J_2 \cos^2\eta) |111\rangle\langle111|+\frac{(J_1-J_2)}{2}\sin2\eta(|000\rangle\langle111|+|111\rangle\langle000|)}{J_1+3J_2}
 \end{equation}
 \end{widetext}
 where $J_1=(1 - \frac{3p_1}{4}) (1 - \frac{3p_2}{4}) (1 - \frac{3p_3}{4})$, $J_2=\frac{1}{64} p_1 p_2 p_3 $ and $p_i$$(i=1,2,3)$ denoting the strength of the three depolarizing channels through which the $1$st, $2$nd and $3$rd qubit of the state are passed respectively. Here $99$th facet inequality detects nonlocality of the noisy version of GGHZ state(\ref{r1}) more efficiently than Svetlichny inequality. For instance, with the strength of depolarization of the three channels being $p_1=0.8,\,p_2=0.7,\,p_3=0.6$ and the state parameter being $\eta=0.69$, $99$th facet inequality(\ref{5}) is violated whereas Svetlichny inequality(\ref{2}) is not. Similar conclusion can be drawn for MS subclass(\ref{10}). For a particular  instance for this subclass, the channels used in the previous example can be considered with the state parameter $\eta=0.69$.
 \subsubsection{Independent amplitude damping}
For practical purposes, analysis of dissipation of energy from a quantum system is an important application of quantum operations. In that context amplitude damping plays a significant role in characterizing the general behavior of some related physical processes. For instance, it may be used to describe the dynamics of an atom which is spontaneously releasing photon, dynamics of an excited spin system tending towards equilibrium with its environment or analyzing state of a photon when subjected to  scattering and attenuation in an
cavity or interferometer. \\
Mathematically, amplitude damping is represented using Krauss operator formalism \cite{NIE,KRA}. When a two qubit state $\varrho$ is passed through an amplitude damping channel, the resulting noisy state is given by
\begin{equation}\label{r2}
    \theta_{AD}(\varrho)=E_0\varrho E_0\dag^+E_1\varrho E_1\dag
\end{equation}
 where $E_0=\left(
              \begin{array}{cc}
                1 & 0 \\
                0 & \sqrt{1-\gamma} \\
              \end{array}
            \right)$
  and $E_1=\left(
              \begin{array}{cc}
                0 & \sqrt{\gamma} \\
                0 & 0 \\
              \end{array}
            \right)$ are the Krauss operators used for representing the amplitude damping operation.  All the three qubits of the state are passed in order through three different amplitude damping channel having varying capacity characterized by the parameters $\gamma_i$$(i=1,2,3)$, precisely $i$th qubit passes through channel characterized by $\gamma_i$$(i=1,2,3)$. In such a situation, noisy version of any state of the GGHZ class is given by:
 \begin{widetext}
  \begin{equation}\label{r3}
\frac{\cos^2\eta |000\rangle\langle000|+(\frac{(D_1+D_2)\sin2\eta}{2})|000\rangle\langle111|+(\frac{D_1\sin2\eta}{2})|111\rangle\langle000|+D_1^2\sin^2\eta|111\rangle\langle111|}{\cos^2\eta+D_1^2\sin^2\eta}
  \end{equation}
 \end{widetext}
 with $D_1=\sqrt{(1-\gamma_1)(1-\gamma_2)(1-\gamma_3)}$ and $D_2=\gamma_1\gamma_2\gamma_3$. As in the case of depolarizing channel, in this case also $99$th facet inequality emerges as a better inequality compared to svetlichny(\ref{2}) to detect nonlocal behavior of any state belonging to this class. For a particular instance, one can consider $1$st, $2$nd and $3$rd qubit of the state $0.995|000\rangle+0.099|111\rangle$ passing through channels with $\gamma_1=0.1,\,\gamma_2= 0.08,\,\gamma_3= 0.09$ respectively. For MS subclass also it can be concluded that $99$th facet inequality(\ref{5}) surpasses Svetlichny inequality(\ref{2}) due to its efficiency to detect genuine nonlocality of the noisy states belonging to this subclass. For a particular numerical example, one may consider the state $\frac{1}{\sqrt{2}}(|000\rangle+0.955|110\rangle+0.296|111\rangle))$ and three amplitude damping channels with $\gamma_1= 0.33,\,\gamma_2= 0.15,\,\gamma_3= 0.09$.

\section{Experimental detection of genuine nonlocality} In recent times  high fidelity tripartite genuine entangled states are generated experimentally(\cite{nat1},\cite{nat2},\cite{nat3},\cite{nat4}). For instance, in \cite{nat2}, an experimental set up is reported where maximally entangled GGHZ state is generated with $86.2\%$ fidelity. So in any such experiment generating genuine tripartite entangled states, the nonlocal behavior of the generated  state can be detected by using suitable  tripartite Bell inequality. As already discussed before, Svetlichny nonlocality lacks proper physical interpretation, to be more precise there exists no clear indication whether signaling is allowed or not while constructing $S_2$ correlations and even if its existence is assumed, then the direction of signaling(one way or both way) cannot be properly interpreted. Due to this sort of ambiguity in definition of $S_2$ correlations, Svetlichny inequality becomes unfit to be used to detect genuine nonlocality of any experimentally generated genuine entangled state. On the contrary, while framing $99$th facet inequality no signaling is assumed. Hence definition of $NS_2$ correlations free from this sort of ambiguity. This in turn points out that the correlation statistics generated while constructing the genuine entangled state experimentally may be used to test the nonlocal nature of the state via the $99$th facet inequality. Apart from the physical point of view, theoretically we have shown that $99$th facet inequality can detect genuine nonlocality of any state belonging to the GGHZ class unlike that of Svetlichny inequality.  This in turn gives advantage for experimental purposes. To be more precise, if the state $0.966|000\rangle+0.259|111\rangle$ is generated experimentally, then Svetlichny inequality fails to detect genuine nonlocality(\ref{8i}) of this genuine tripartite entangled state($\tau=0.25<\frac{1}{3}$)whereas $99$th facet can detect its nonlocal behavior from the correlation statistics(\ref{8}).
\section{Family of some higher rank mixed states} For a three-qubit mixed state,  $\rho=\sum_{i}p_i\rho_i$  analytical expressions of three-tangle are given in terms of convex roof: $\tau(\rho)=\min_{p_i}\sum_ip_i\tau(\rho_i)$. So far the analytical expressions for the three-tangle ($\tau$) for all states of rank-2 \cite{LOH,ELT}, rank-3, 4 \cite{EJU,JUN} and rank-n where $n=5,6,7,8$ \cite{SJH} are reported. Mixture of GHZ and W states and that of GHZ, W and flipped W states are mixed states of rank-2 and rank-3 respectively:
\begin{eqnarray}
\rho_2&=&  p\,|GHZ\rangle\langle GHZ|\,+\,(1-p)\,|W\rangle\langle W| \\
\rho_3^k &=& p|GHZ\rangle\langle GHZ|+q|W\rangle\langle W|+(1-p-q)|\widetilde{W}\rangle\langle \widetilde{W}|
\end{eqnarray}
where $|GHZ\rangle=\frac{|000\rangle+|111\rangle}{\sqrt{2}}$, $|W\rangle=\frac{|001\rangle+|010\rangle+|100\rangle}{\sqrt{3}}$, $|\widetilde{W}\rangle=\frac{|011\rangle+|110\rangle+|101\rangle}{\sqrt{3}}.$ and $q=\frac{1-p}{k}$, $k$ being any positive integer. For $k=1$, $\rho_3^k=\rho_2$. The study of three-tangle ($\tau$) of these mixed states focusses on the characterization of genuine tripartite entanglement. But none of these studies demonstrates the nonlocal behavior of these states. The range of $p$ for which rank-2 and rank-3 (for $k\,=\,2,3,10$) mixed states exhibit nonlocality via violation of $99$-th facet (\ref{5}) and Svetlichny inequality(\ref{2}) is summarised in  Table \ref{Table1}.
\begin{table}[htp]
	\begin{center}
		\begin{tabular}{|c|c|c|c|}
			\hline
			State & $\tau>0$ & Violation of Eq.(\ref{5})&Violation of Eq.(\ref{2})\\
			\hline
			$\rho_2$ &  $p\geq0.6268$&$p\geq0.811876 $&$p\geq0.707109$\\
			\hline
			$\rho_3^2$ & $p\geq0.75$ & $p\geq0.819964$&$p\geq0.70719$ \\
			\hline
			$\rho_3^3$ &$p\geq0.7452$&$p\geq0.818825$&$p\geq0.707109$\\
			\hline
			$\rho_3^{10}$ &$p\geq0.7452$&$p\geq0.814789$&$p\geq0.707109$\\
			\hline
		\end{tabular}\\
	\end{center}
	\caption{The table gives a comparison between the bounds for violation of Svetlichny inequality and $99$th facet inequality along with that of the measure of tripartite entanglement($\tau$) for some of the mixed states $\rho_2$, $\rho_3^i$(i=2,3,10). }
	\label{Table1}
\end{table}

As pointed out in \cite{SJH}, the families  of high rank (n\,=\,5,...,8) mixed states are interesting for practical purposes as their tripartite entanglement features would reveal further relations between the quantum phase transitions and quantum entanglement just like  concurrence of mixed two-qubit states which have been applied in the study of quantum phase transitions. Besides, as $\tau$, being a measure of genuine tripartite entanglement, is capable of giving certain information in the scheme of quantum copy machine or three-party quantum teleportation \cite{KAR}. So a systematic study dealing with the connection of this measure of genuine entanglement with that of violation of a Bell-type inequality (facet of a local polytope) can be useful in related research works.  The families of rank $n$ ($4,...8$) are \cite{JUN,SJH}:
\begin{widetext}
\begin{eqnarray}
 \rho_4 &=& p|\Lambda,1+\rangle\langle \Lambda,1+|+\frac{(1-p)}{3}\Pi\,\,(\textmd{n=4})\\
  \rho_5 &=& p|\Lambda,1+\rangle\langle \Lambda,1+|+\frac{(1-p)}{10}(|\Lambda,1-\rangle\langle \Lambda,1-|+3\Pi)\,\,(\textmd{n=5}) \\
  \rho_6 &=& p|\Lambda,2-\rangle\langle \Lambda,2-|+\frac{(1-p)}{11}(\Omega+3\Pi)\,\,(\textmd{n=6}) \\
 \rho_7 &=& p|\Lambda,3-\rangle\langle \Lambda,3-|+\frac{(1-p)}{34}(|\Lambda,2-\rangle\langle \Lambda,2-|+3\Omega+9\Pi)\,\,(\textmd{n=7})\\
\rho_8 &=& p|\Lambda,4-\rangle\langle \Lambda,4-|+\frac{(1-p)}{35}(|\Lambda,2-\rangle\langle \Lambda,2-|+|\Lambda,3-\rangle\langle \Lambda,3-|+3\Omega+9\Pi)\,\,(\textmd{n=8})
\end{eqnarray}
where $|\Lambda,1\pm\rangle=\frac{|000\rangle\pm|111\rangle}{\sqrt{2}}, ~~ |\Lambda,2\pm\rangle=\frac{|110\rangle\pm|001\rangle}{\sqrt{2}}, ~~|\Lambda,3\pm\rangle=\frac{|101\rangle\pm|010\rangle}{\sqrt{2}}, ~~|\Lambda,4\pm\rangle=\frac{|011\rangle\pm|100\rangle}{\sqrt{2}}, ~~\Omega\,=\,|\Lambda,1+\rangle\langle\Lambda,1+| + |\Lambda,1-\rangle\langle\Lambda,1-|$ and $\Pi\,=\,|\Lambda,2+\rangle\langle \Lambda,2+|+|\Lambda,3+\rangle\langle \Lambda,3+|+|\Lambda,4+\rangle\langle \Lambda,4+|$.
\end{widetext}
The closed form of the bounds ($B_6$ and $B_7$) of $99$-th facet inequality for $\rho_4$ and $\rho_5$ are  respectively:
  \begin{eqnarray}
      B_6 &=& \frac{ 2 \sqrt{16 p^2 - 8 p + 10} + |1 -4 p|}{3} \\
    B_7 &=&\frac{ 2 \sqrt{37 p^2 - 4 p + 17} + |1- 6 p|}{5}.
  \end{eqnarray}
$\rho_4$ violates Eq.(\ref{5}) for $p\geq0.726$ and Svetlichny inequality (\ref{2}) for $p\geq0.72,$ whereas it has a nonzero tangle for $p\geq0.75$ \cite{JUN}. As genuine multipartite nonlocal nature of the correlations produced by a state is a signature of existence of genuine multipartite entanglement of the state, the $185$-th facet (i.e., Svetlichny inequality (\ref{2})) can be regarded as a better measure of genuine tripartite entanglement of $\rho_5$ than the three tangle($\tau$).  A similar sort of results holds for $\rho_5$ which violates Eq.(\ref{2}) and Eq.(\ref{5}) for $p\geq0.710858$ and $p\geq0.729157$ respectively, but has $\tau>0$ for $p\geq0.737$ \cite{SJH}. Hence for $p\in[0.710858,\,0.737]$ violation of a facet inequality of $NS_2$ local polytope serves as a better measure for detecting genuine tripartite entanglement compared to three tangle. However, a comparative  study of the range of $p$ for non zero three tangle and that of $99$-th (\ref{5}) and $185$-th (\ref{2}) facet violation for the other three families of higher rank mixed states ($n=6,7,8$) does not yield analogous result and so these facet inequalities can only be used to demonstrate the genuine tripartite nonlocality of the correlations produced in a system using these families of mixed states upto projective measurements. The results obtained are summarized as follows(Table \ref{Table2}):
\begin{widetext}
	\begin{table}[htp]
		\begin{center}
			\begin{tabular}{|c|c|c|c|c|}
				\hline
				State & Bound for Eq.(\ref{5})&$\tau>0$ & Violation of Eq.(\ref{5})&Violation of Eq.(\ref{2})\\
				\hline
				$\rho_6$ &$\frac{\sqrt{(1 + 10 p)^2 + (6 (1 - p) + |3 - 14 p|)^2} +
					|(12 p - 1)|}{11}$ & $p\geq0.2143$&$p\geq0.756458 $&$p\geq0.765134$\\
				\hline
				$\rho_7$ &$\sqrt{(-.11765 + 1.11765 p)^2+B^2}+\sqrt{(.11765 + 0.8824 p)^2+B^2}$&\,&\,\\
				\,&$+0.0588 + 0.9412 p\,,\, B=\frac{1 - p}{2} + |0.26470 - 1.26470 p|$ & $p\geq0.2062$ & $p\geq0.759185$&$p\geq0.76444$ \\
				\hline
				$\rho_8$ &$\sqrt{(0.0857 + 0.9143 p)^2 +C^2}+\sqrt{(-0.1428 + 1.1429 p)^2+C^2}$&\,&\,\\
				\,&$+0.0857 + 0.9142 p\,,\,C=0.4572 (1 - p) + |.2571 - 1.2571 p|$&$p\geq0.2490$&$p\geq0.75843$&$p\geq0.763645$\\
				\hline
			\end{tabular}\\
			\caption{The table gives a comparison between the bounds for violation of Svetlichny inequality and $99$th facet inequality along with that of the measure of tripartite entanglement($\tau$) for some of the higher rank mixed states $\rho_6$, $\rho_7$ and $\rho_8$. }
			\label{Table2}
		\end{center}
	\end{table}
\end{widetext}
\section{Conclusion:}  In conclusion, the above systematic study exploiting the relation between genuine tripartite entanglement measure and that of genuine tripartite nonlocality reveals the efficiency of a facet ($99$-th) of $NS_2$ local polytope as a tripartite Bell-type inequality over Svetlichny inequality (which is reported as the $185$-th facet of $NS_2$ local polytope) for detecting genuine three way nonlocality of GGHZ and a subclass $S$ of extended GHZ class of pure tripartite genuine entangled states. This in turn helps to prove the conjecture made by Bancal in(\cite{BAL}) for the GGHZ class of states. Apart from the obvious changes in the analytical expressions of the bounds of $99$-th facet inequality from that of Svetlichny inequality (\ref{2}) for the corresponding class of states, it is interesting to note that the rectification (made by Bancal in (\cite{BAL})), of local-nonlocal hybrid correlations (\ref{1}) and thereby giving a more physically well motivated definition of the same (\ref{1},\ref{4i},\ref{4ii}),  contributes in developing more efficient measures for demonstrating three way nonlocality. So far this sort of analysis has been reported only in continuous variable systems of Gaussian states (\cite{ADE}). A comparative study between discord monogamy score and violation of these two facet inequalities clearly points out that from the perspective of revealing quantumness in a multipartite system, $99$-th facet inequality (\ref{5}) surpasses Svetlichnhy inequality (\ref{2}). Besides, in the context of analyzing the interplay between mixed state genuine entanglement and the nonlocality of the corresponding correlations emerging in a quantum system,  Eq.(\ref{5}) gives advantage over Svetlichny inequality (\ref{2}) in many cases such as lowering down the local visibility threshold $V$ of a subclass of noisy GGHZ and subclass $S$ of extended GHZ states and also for some family of higher rank mixed states. Here it is worth mentioning that for some family of mixed states of rank-4 and rank-5, these facet inequalities (both \ref{2} and \ref{5}) can even be used as a better measure of genuine tripartite entanglement over the usual measure of three tangle ($\tau$). It will be interesting to explore further in this approach that whether one can develop a more efficient measure of detecting three way nonlocality (both mixed and pure) and also if possible a better measure of genuine entanglement for other family of mixed states by choosing a suitable facet of the $NS_2$ local polytope and also by performing more general measurements which in turn may lead to prove the conjecture made by Bancal et.al. (\cite{BAL}). Apart from this, there are many other future directions. The analysis made here can be used in related physical experiments. One may also explore the relation between tripartite information content of GGHZ class \cite{LND} and the corresponding nonlocal correlations. There exists close correspondence between Bell inequality violation and nonlocal games \cite{SIL}. So from that perspective it we will be interesting to develop a detailed study focussing on the efficiency of GGHZ and subclass $S$ of extended GHZ class in various protocols based on nonlocal games \cite{CVE}.

\textit{Acknowledgement:}
The authors acknowledge fruitful discussions with Mr. A.Sen and Ms. S.Karmakar. The author KM acknowledges financial support from University Grants Commission(UGC), New Delhi and the author D. Sarkar acknowledges the support from DST, India.

 \section{Appendix A}
In order to obtain the bound given in (\ref{7}), we consider the following measurements: $X = \vec{x}.\vec{\sigma_1} $ or, $\acute{X} = \vec{\acute{x}}.\vec{\sigma_1}$ on qubit 1, $Y = \vec{y}.\vec{\sigma_2} $ or, $\acute{Y} = \vec{\acute{y}}.\vec{\sigma_2}$ on qubit 2, and $Z = \vec{z}.\vec{\sigma_3} $ or, $\acute{Z} = \vec{\acute{z}}.\vec{\sigma_3}$ on qubit 3, where $\vec{x},\vec{\acute{x}},\vec{y},\vec{\acute{y}}$ and $\vec{z},\vec{\acute{z}}$ are unit vectors and $\sigma_i$ are the spin projection operators that can be written in terms of the Pauli matrices. Representing the unit vectors in spherical coordinate, we have, $\vec{x} = (\sin\theta a_0 \cos\phi a_0, \sin\theta a_0 \sin\phi a_0, \cos\theta a_0), ~~\vec{y} = (\sin\alpha b_0 \cos\beta b_0, \sin\alpha b_0 \sin\beta b_0, \cos\alpha b_0) $ and $\vec{z} = (\sin\zeta c_0 \cos\eta c_0, \sin\zeta c_0 \sin\eta c_0, \cos\zeta c_0) $ and similarly, we define, $\vec{\acute{x}},\vec{\acute{y}}$ and $\vec{\acute{z}}$ by replacing $0$ in the indices by $1$. Then the expectation value of the operator $NS$ (\ref{5}) with respect to the state $|\varphi_{GGHZ}\rangle$ gives
$$  NS(|\varphi_{GGHZ}\rangle) = \cos(\theta a_0)\cos(\alpha b_0)(\cos(\zeta c_0) \cos(2\eta) -$$

$\cos(\zeta c_1)\cos (2\eta))+  \sin(\theta a_0)\sin(\alpha b_0)(\sin(\zeta c_0)\cos(\beta b0 + \eta c_0 +\phi a_0)\sin (2\eta)- \sin(\zeta c_1)\cos(\beta b0 + \eta c_1 +\phi a_0)\sin 2\eta)$
 \begin{equation}\label{A1}
  + \cos(\zeta c_0)\cos(\alpha b_1) + \cos(\alpha b_1)\cos(\theta a_1)+ \cos(\zeta c_1)\cos(\theta a_1).
 \end{equation}
Hence in order to get a maximum value of the $NS$ operator defined in (\ref{5}), we have to perform a maximization over $12$ measurement angles. We first find the global maximum of $NS(|\varphi_{GGHZ}\rangle)$ with respect to $\theta a_0$ and $\alpha b_0$. We begin with by finding all critical points of $NS(|\varphi_{GGHZ}\rangle)$ inside the region $R=[0 , 2\pi]\times [0 , 2\pi]$ which are namely $(0,0)$, $(\frac{\pi}{2}, -\frac{\pi}{2})$,$(-\frac{\pi}{2}, \frac{\pi}{2})$ and $(\frac{\pi}{2}, \frac{\pi}{2})$. Among all these critical points the point $(\frac{\pi}{2}, \frac{\pi}{2})$ gives the global maximum of $NS(|\varphi_{GGHZ}\rangle)$. Thus, we have,
 $$ NS(|\varphi_{GGHZ}\rangle) \leq \sin(\zeta c_0)\cos(\beta b0 + \eta c_0$$
 $+\phi a_0)\sin (2\eta) - \sin(\zeta c_1)\cos(\beta b0 + \eta c_1 +\phi a_0)\sin (2\eta)+ $
\begin{equation}\label{A2}
    \cos(\zeta c_0)\cos(\alpha b_1)+\cos(\zeta c_1)\cos(\theta a_1) + \cos(\alpha b_1)\cos(\theta a_1)
\end{equation}
which when maximized with respect to $\zeta c_0$ and $\zeta c_1$, gives
$$NS(|\varphi_{GGHZ}\rangle) \leq \sqrt{cos^2(\alpha b_1) + \sin^2 (2\eta)} $$
\begin{equation}\label{A3}
+ \sqrt{cos^2(\theta a_1) + \sin^2 (2\eta)} + \cos(\alpha b_1)\cos(\theta a_1).
\end{equation}
The last inequality is obtained by using the inequality $x\cos\theta + y\sin\theta \leq \sqrt{x^2 + y^2}$ (the equality holds for $\tan\theta = \frac{y}{x}$) and by letting $\cos^2(\beta b0 +\eta c_0  +\phi a_0)=1, ~~\cos^2(\beta b0 + \eta c_1 +\phi a_0) = 1.$ Again by classifying all the critical points, it is easy to check that the expression in Eq.(\ref{A3}) gives the global maximum with respect to $\alpha b_1$ and $\theta a_1$ for the critical point $(0,0)$. Therefore, Eq.(\ref{A3}) reduces to:
\begin{equation}\label{A4}
 NS(|\varphi_{GGHZ}\rangle)\leq 1 + 2\sqrt{1 + \sin^2 2\eta}
\end{equation}
as stated in Eq.(\ref{7}).

\section{Appendix B}
For extended GHZ states, the expression of $NS$ (\ref{5}) is:
\begin{widetext}
$NS(|\chi\rangle)=A_0((1-2\lambda_4^2)Y_1-2\lambda_3 \lambda_4 \sin(\zeta c1)\cos(\eta c1))+D_0(2\lambda_0\lambda_3 Y_1B_0+2\lambda_0\lambda_4
\sin(\zeta c0)C_0-2\lambda_0\lambda_4 \sin(\zeta c1)C_1)$
\begin{equation}\label{B1}
+Y_0(1-2\lambda_3^2)-2\lambda_3\lambda_4\sin(\zeta c0)\cos(\eta c0)\cos(\alpha b1)-2\lambda_3\lambda_4\sin(\zeta c1)\cos(\eta c1)\cos(\theta a1)+A_1+2\lambda_0\lambda_3 B_1D_1.
\end{equation}
\end{widetext}
where $X_i=\cos(\zeta ci)$, $A_i$ $=\cos(\alpha bi)\cos(\theta ai)$, $B_i=\cos(\beta bi+\phi ai)$, $C_i=\cos(\beta b0+\phi a0+\eta ci)$, $D_i$ $=\sin(\alpha b0)\sin(\theta a0)(i=0,1)$ and $X_0+X_1=Y_0$, $X_0-X_1=Y_1$

Considering above expression as a function of $\sin(\theta a0)$ and $\sin(\alpha b0),$  we get the critical points: $(0,\,0)$, $(1,\,1)$, $(1,\,-1)$, $(-1,\,1)$, $(-1,\,-1).$ The global maximum of this function gives the required bound.

\textit{Case1}: For $\sin(\theta a0)\,=\,0$ and $\sin(\alpha b0)\,=\,0$
\begin{widetext}
$$NS(|\chi\rangle)=X_0(T_1\cos(\alpha b1)+T_2)+\sin(\zeta c0)(-2T_3\cos(\eta c0)\cos(\alpha b1)+2T_3\cos(\eta c0))X_1(T_1\cos(\theta a1)$$ $$-T_2)+\sin(\zeta c1)(-2T_3\cos(\eta c1)\cos(\theta a1)-2T_3\cos(\eta c1))+A_1+2\lambda_0\lambda_3B_1D_1$$
$$\leq\sqrt{\cos^2(\alpha b1)(T_1^2+(2T_3\cos(\eta c0))^2)+2(T_1T_2-4(T_3)^2)\cos(\alpha b1)+(2T_3\cos(\eta c0))^2}+$$ $$\sqrt{\cos^2(\theta a1)(T_1^2+(2T_3\cos(\eta c1))^2)+2(T_1T_2-4(T_3)^2)\cos(\theta a1)+(2T_3\cos(\eta c1))^2}+A_1+2\lambda_0\lambda_3B_1D_1.$$
\end{widetext}
where $T_1$=$(1-2\lambda_3^2)$, $T_2$=$(1-2\lambda_4^2)$ and $T_3$=$\lambda_3\lambda_4$
The above expression can be considered as a function of $\cos(\alpha b1)$ and $\cos(\theta a1)$ which is maximum for $\cos(\alpha b1)=0$ and $\cos(\theta a1)=0$. Hence the bound gets modified as
$$\sqrt{T_1^2+T_2^2+2T_1T_2}+\sqrt{T_1^2+T_2^2-2T_1T_2+(4T_3\cos(\eta c1))^2}+1.$$
which is maximum for $\eta c1\,=\,0$. So, we get the bound as, $$\sqrt{4\lambda_0^4}+\sqrt{4(\lambda_3^2-\lambda_4^2)^2+16\lambda_3^2\lambda_4^2}+1$$ which on further simplification gives the value $3.$\\

\textit{Case2:} For $\sin(\theta a0)\,=\,1$ and $\sin(\alpha b0)\,=\,-1$
\begin{widetext}
$$NS(|\chi\rangle)\,=X_0(-2\lambda_0\lambda_3 B_0+T_1\cos(\alpha b1))
+\sin(\zeta c0)(-2\lambda_0\lambda_4C_0-2T_3\cos(\eta c0)\cos(\alpha b1))+\cos(\zeta c1)(-2\lambda_0\lambda_3 B_0+T_1\cos(\theta a1))$$
$$+\sin(\zeta c1)(-2\lambda_0\lambda_4C_1-2T_3\cos(\eta c1)\cos(\theta a1))+A_1+2\lambda_0\lambda_3B_1D_1$$
$$\leq \sqrt{(2\lambda_0\lambda_3 B_0-T_1\cos(\alpha b1))^2+(-2\lambda_0\lambda_4C_0-2T_3\cos(\eta c0)\cos(\alpha b1))^2}$$
$$+\sqrt{(2\lambda_0\lambda_3 B_0+T_1\cos(\theta a1))^2+(2\lambda_0\lambda_4C_1-2T_3\cos(\eta c1)\cos(\theta a1))^2}$$ $\,=\,G(\cos(\alpha b1),\,\cos(\theta a1))$
\end{widetext}

which is maximum for $\cos(\alpha b1)=0$ and $\cos(\theta a1)=0$. Thus, the bound modified as:
$$\sqrt{(2\lambda_0\lambda_3 B_0-T_1)^2+(2\lambda_0\lambda_4C_0+2T_3\cos(\eta c0))^2}$$
$$+\sqrt{(2\lambda_0\lambda_3 B_0+T_1)^2+(2\lambda_0\lambda_4C_1-2T_3\cos(\eta c1))^2}.$$
Putting $\beta b0=\phi a0=0$, and then maximizing over $\eta c0$ and $\eta c1$(maximum for $\eta c0=\eta c1=0$ ), we get the bound:
 \begin{widetext}
 $$\sqrt{(2\lambda_0\lambda_3 -T_1)^2+(2\lambda_0\lambda_4+2T_3)^2}+\sqrt{(2\lambda_0\lambda_3 +T_1)^2+(2\lambda_0\lambda_4-2T_3)^2}$$
\end{widetext}
 This on further simplification gives
$$ 1+\sqrt{1+4\lambda_0^2\lambda_4^2+4\lambda_0\lambda_3(1-2\lambda_0^2)}$$
$$+\sqrt{1+4\lambda_0^2\lambda_4^2-4\lambda_0\lambda_3(1-2\lambda_0^2)}.$$

So, $f(1,\,-1)=1+\sqrt{1+4\lambda_0^2\lambda_4^2+4\lambda_0\lambda_3(1-2\lambda_0^2)}+\sqrt{1+4\lambda_0^2\lambda_4^2-4\lambda_0\lambda_3(1-2\lambda_0^2)}.$

Similarly, it can be shown that $f(-1,\,1)=f(1,\,1)=f(-1,\,-1)=f(1,\,-1)$.

Using $\tau=4\lambda_0^2\lambda_4^2$ and $C_{12}^2=4\lambda_0^2\lambda_3^2$ we get: $(1-2\lambda_0^2)^2\,=\,1-(\tau+C_{12}^2)$ .

Hence $$f(1,\,-1)=1\,+\,\sqrt{1+\tau+2\sqrt{C_{12}^2(1-(\tau+C_{12}^2))}}$$
$~~~~~~~~~~~+\sqrt{1+\tau-2\sqrt{C_{12}^2(1-(\tau+C_{12}^2))}}.$

Clearly,

$f(1,-1)>3$ if $\sqrt{1+\tau+2\sqrt{C_{12}^2(1-(\tau+C_{12}^2))}}$ $$+\sqrt{1+\tau-2\sqrt{C_{12}^2(1-(\tau+C_{12}^2))}}>2$$ which on simplification gives, $f(1,-1)>3$ if $\tau>C_{12}^2-\tau C_{12}^2-C_{12}^4,$ i.e., $\tau>\frac{C_{12}^2(1-C_{12}^2)}{1+C_{12}^2}.$ So, the global maximum of $f(\sin(\theta a0),\,\sin(\alpha b0))$ is given by $$\max\{3,\,1\,+\,\sqrt{1+\tau+2\sqrt{C_{12}^2(1-(\tau+C_{12}^2))}}$$
$+\sqrt{1+\tau-2\sqrt{C_{12}^2(1-(\tau+C_{12}^2))}}\}$

Thus we get the required bound Eq.(\ref{13}). For MS subclass, $\lambda_0=\frac{1}{\sqrt{2}}$, $\lambda_3=\frac{\cos(\eta)}{\sqrt{2}}$ and $\lambda_4=\frac{\sin(\eta)}{\sqrt{2}}$. So $\tau=\sin^2(\eta)$ and $C_{12}^2=\cos^2(\eta)$. Hence the bound becomes:
 $$1+\sqrt{1+\sin^2(\eta)+\cos^2(\eta)(1-\sin^2(\eta)-\cos^2(\eta))}$$
$+\sqrt{1+\sin^2(\eta)-\cos^2(\eta)(1-\sin^2(\eta)-\cos^2(\eta))}$

$~~~~~~~~=1+2\sqrt{1+\sin^2(\eta)}$

which is the bound (\ref{11}) for Maximal Slice states.
\section{Appendix C}
\textit{Calculation of monogamy score $\delta_D$ of GGHZ class and subclass $S$ of extended GHZ class of states:}
The mutual information, for a quantum state $\rho_{AB}$, is given by \cite{ZUR,NCE,SCH,GRO,BAN1}:
\begin{equation}\label{c1}
    \mathcal{I}(\rho_{AB})=\mathcal{S}(\rho_{A})+\mathcal{S}(\rho_{B})-\mathcal{S}(\rho_{AB}).
\end{equation}
where $\mathcal{S}(\rho)$ denotes the Von Neumann entropy of the state $\rho.$
The classical mutual information for a bipartite quantum state $\rho_{AB}$:
\begin{equation}\label{c2}
    \mathcal{J}(\rho_{AB})=\mathcal{S}(\rho_{A})-\mathcal{S}(\rho_{A/B}).
\end{equation}
$\mathcal{S}(\rho_{A/B})$ is the quantum conditional entropy given by:
\begin{equation}\label{c3}
    \mathcal{S}(\rho_{A/B})=min_{\Pi_i^B}\sum_{i}p_i\mathcal{S}(\rho_{A/i})
\end{equation}
where the minimization is over all rank-1 measurements, {$\Pi_i^B$} performed over subsystem $B$. Here $p_i$ is the corresponding probability for obtaining the outcome $i$ and $\rho_{A/i}$ denotes the corresponding post measurement state of the subsystem $A.$ The quantum discord \cite{OLI,MOD} of a state $\rho_{AB}$ is given  by:
\begin{equation}\label{c4}
    \mathcal{D}(\rho_{AB})=\mathcal{I}(\rho_{AB})-\mathcal{J}(\rho_{AB}).
\end{equation}
For GGHZ class of states, $\rho_{ABC}\,=\, |\varphi_{GGHZ}\rangle\langle \varphi_{GGHZ}|.$ $\rho_{AB}\,=\,\rho_{AC}\,=\,\cos^2(\eta) |00\rangle\langle00|+\sin^2(\eta) |11\rangle\langle11|$. Now both of $\rho_{AB}$ and $\rho_{AC}$ are classical quantum states (i.e., they are of the form: $\sum_{p_k}\rho_k\otimes|k\rangle\langle k|$). Hence $D(\rho_{AB})\,=\,D(\rho_{AC})\,=\,0$ (discord of a classical state vanishes). $\rho_{A:BC}$ is a bipartite (with $A$ as one party and $B$ and $C$ together taken as another party) pure state. As quantum discord of a bipartite pure state is equal to its measure of entanglement, $D(\rho_{A:BC})$ is given by the Von Neumann entropy of the corresponding reduced states. Hence $$D(\rho_{A:BC})\,=\,\mathcal{S}(\rho_{A})$$
$$=\,-(\cos^2\eta\log_2(\cos^2\eta) + \sin^2\eta\log_2(\sin^2\eta)).$$ This gives the the discord monogamy score of $|\varphi_GHZ\rangle.$
For subclass $S$ of extended GHZ class of states $\rho_{ABC}\,=\, |\chi\rangle\langle \chi|.$ $$\rho_{AB}\,=\,\lambda_0^2|00\rangle\langle00|+(1-\lambda_0^2)|11\rangle\langle11|+\lambda_3\lambda_0(|00\rangle\langle11|+|11\rangle\langle00|).$$ In order to calculate the discord of a state one has to calculate Eq.(\ref{c3}) which is hard to compute for obvious reasons. However $\rho_{AB}$ belongs to the class of X states for which an algorithm calculating discord is reported in \cite{ZHN}. Following the technique used there we get
\begin{widetext}
$$\frac{-\ln4(\lambda_0^2\ln(\lambda_0^2)\,+\,(1-\lambda_0^2)\ln(1-\lambda_0^2))\,+\,\ln(2)(1+r\log_2(\frac{1+r}{2})\,+\,1-r\log_2(\frac{1-r}{2}))}{\ln(2)\ln(4)}$$ where $r\,=\,\sqrt{1+4 \lambda_0^4 + 4 \lambda_0^2 (-1 + \lambda_3^2)}$.
\end{widetext}
Now $\rho_{AC}\,=\,\lambda_0^2|00\rangle\langle00|\,+\,\lambda_4^2|11\rangle\langle11|\,+\,\lambda_3^2|10\rangle\langle10|\,+\,T_3(|11\rangle\langle10|\,+\,|10\rangle\langle11|).$ It is a classical quantum state of the form $|k\rangle\langle k|\otimes\sum_{p_k}\rho_k$. Hence $D(\rho_{AC})=0$. Here $\rho_{A:BC}$ is a bipartite (with $A$ as one party and $B$ and $C$ together taken as another party) pure state. Hence $D(\rho_{A:BC})\,=\,-(\lambda_0^2\log_2(\lambda_0^2)\,+\,(1-\lambda_0^2)\log_2(1-\lambda_0^2))$. On further simplification and using $\tau=4\lambda_0^2\lambda_4^2$ and $C_{12}^2=4\lambda_0^2\lambda_3^2$ in Eq.(\ref{6ii}), we get the discord monogamy score (\ref{13i}).
\section{Appendix D}
\textit{Relation between $\tau$ and $C_{12}^2$ for states belonging to extended GHZ class of states characterized by $\lambda_0=\frac{1}{\sqrt{2}}:$}
As $\lambda_0=\frac{1}{\sqrt{2}}$,
\begin{equation}\label{d1}
    \tau=2\lambda_4^2
\end{equation}
 and
 \begin{equation}\label{d2}
    C_{12}^2=2\lambda_3^2
 \end{equation}
 Hence
\begin{equation}\label{d3}
    \tau=1-C_{12}^2.
\end{equation}
Hence $\tau>\frac{1-C_{12}^2}{2}$ which in turn implies that for $\lambda_0=\frac{1}{\sqrt{2}}$ Svetlichny inequality(\ref{2}) is always violated. Now we focus on the range of $\tau$ occurring in the bound of the $99$th facet inequality(\ref{5}). We claim that $\tau$ always lies inside the interval $(\frac{C_{12}^2(1-C_{12}^2)}{1+C_{12}^2},\,1].$
If possible let $\tau$ lies outside the interval $(\frac{C_{12}^2(1-C_{12}^2)}{1+C_{12}^2},\,1]$, i.e. there exists some value $\tau$ lying outside this interval. This gives $\tau<\frac{C_{12}^2(1-C_{12}^2)}{1+C_{12}^2}$ which along with Eqs.(\ref{d1}, \ref{d2}) implies $1<0.$ This proves our claim which in turn implies for $\lambda_0=\frac{1}{\sqrt{2}}$, $99$ facet inequality(\ref{5}) is violated for any positive value of $\tau$ and hence by any tripartite pure state belonging to the subclass $S$ and characterized by $\lambda_0=\frac{1}{\sqrt{2}}$.
\end{document}